\documentclass[10pt]{article}
\usepackage{hyperref}
 \usepackage{footnote}
\makesavenoteenv{minipage}

\title{Anti-FFBRST Transformations for the BLG Theory in Presence of a  Boundary}
\author{
{  {\normalsize Mir Faizal }$^{a}
$\thanks{f2mir@uwaterloo.ca}},
{  {\normalsize Sudhaker Upadhyay}$^{b}$\thanks{   sudhakerupadhyay@gmail.com }},
{  {\normalsize Bhabani P Mandal}$^{c}$\thanks{ bhabani.mandal@gmail.com}}\\
$^{a}${\normalsize Department of 
Physics and Astronomy, University of Waterloo, }\\
{\normalsize Waterloo, Ontario N2L 3G1, Canada}\\
$^{b}${\normalsize Department of Physics, Indian Institute of Technology Kanpur, Kanpur 208016, India } \\
$^{c}${\normalsize  Department of Physics, Banaras Hindu University,   
Varanasi 221005, India}\\
\\[0.3cm]
}
\date{}

\begin{document}

\maketitle

\begin{abstract}

In this paper we will analyse the 
 anti-BRST symmetries of Bagger-Lambert-Gustavsson (BLG) theory 
in presence of a boundary.
We will analyses these symmetries in both linear and non-linear gauges. We will also  derive the
finite field version of the
 anti-BRST transformations for the 
BLG theory in presence of a boundary. These finite field transformations will be used to 
relate generating functional  
in  linear gauge to the generating functional in the non-linear gauge. 

\end{abstract}
\section{Introduction} 
  According to the $AdS/CFT$ correspondence the superconformal field theory describing 
  multiple M2-branes is dual to
the eleven dimensional supergravity  on $AdS_4 \times S^7$.
 Now apart from a constant closed 7-form on $S^7$,
$AdS_4 \times S^7 \sim [SO(2,3)/ SO (1, 3)]\times  [SO(8)/ SO(7)] \subset OSp(8|4)/[SO(1,3) \times SO(7)]$ \cite{aaaa}, 
  so the superconformal field theory  dual to the eleven dimensional supergravity has $\mathcal{N} = 8$ supersymmetry. 
  This 
  is because for this dual superconformal field theory
  $OSp(8|4)$ gets realized as $\mathcal{N} = 8$ supersymmetry. 
  This superconformal field theory contains eight gauge valued scalar fields which originate from the transversal coordinates 
  of M2-branes. It also contains sixteen  physical fermions. 
 Furthermore,  it  only has sixteen on shell degrees of freedom and so the gauge fields  cannot 
have any contribution to the on shell degrees of freedom. These requirements are met by a theory called 
the Bagger-Lambert-Gustavsson (BLG) theory \cite{1}-\cite{5}. This theory 
is a three dimensional superconformal field theory with 
$\mathcal{N} = 8$ supersymmetry and  $SO(8)$ R-symmetry.

 The BLG theory has gauge symmetry and is valued in a  Lie $3$-algebra rather than a conventional Lie algebra.  
 The only known example of a  Lie $3$-algebra with fully antisymmetric structure constants is $SO(4)$ and it has
 not been possible to increase the rank of the gauge group.
 It is possible to decompose the gauge group  $SO(4)$ into $SU (2) \times SU (2)$, 
 by complexifing the matter fields. The gauge sector of the BLG theory now comprises of two 
 Chern-Simons theories with levels $\pm k$ and the matter
 fields exist in the bi-fundamental representation 
 of the gauge group $SU (2) \times SU (2)$. However, the gauge symmetry is now produced  by ordinary Lie algebra 
 rather than a Lie $3$-algebra. As the gauge group of the BLG theory  is
 $SU(2) \times SU(2)$, it only represents two M2-branes. It has been possible to extend the gauge group 
 to  $U(N) \times U(N)$, and the resultant theory is called Aharony-Bergman-Jafferis-Maldacena (ABJM) 
theory \cite{abjm}.    The ABJM theory only has $\mathcal{N} =6$ supersymmetry, and 
this supersymmetry   gets enhanced to 
$\mathcal{N} =8$ supersymmetry 
only for Chern-Simons levels, $k = 1, 2$ \cite{abjm2}.   
Thus, the ABJM theory only has
$\mathcal{N} =6$ supersymmetry instead of $\mathcal{N} =8$ supersymmetry, which M2-branes are expected to have from the 
$AdS/CFT$ correspondence.
However, for two M2-branes ABJM theory coincides with the BLG theory and thus has $\mathcal{N} =8$ supersymmetry.

Just as strings can end on D-branes in string theory, 
M2-branes can end on other objects in M-theory. 
Thus, M2-branes can end on M5-branes, M9-branes and gravitational waves.
 In fact, 
a form of non-commutative string theory on the M5-brane world-volume has been studied by 
analyzing  the action of a single open M2-branes ending on it
\cite{NCS1}-\cite{NCS3}.
Furthermore, the BLG theory  has  been used  for constructing 
a novel quantum geometry on the M5-brane world-volume by analysing
a system of M2-branes ending on a M5-brane with a constant $C$-field
\cite{d12}. The BLG action with Nambu-Poisson 3-bracket has been 
identified 
as the M5-brane action with a
large world-volume C-field \cite{M5BLG}. As the BLG theory 
 has been used for analyzing a system of M2-branes ending on a M5-brane, it is
is important to study the BLG theory in presence of a boundary. 

In a supersymmetric theory, the presence of a boundary  breaks  the supersymmetry.  This is because the 
boundary
obviously breaks translational symmetry and since supersymmetry closes on translations,
 it is inevitable that the presence of boundary will also break the supersymmetry.
However, half of the the supersymmetry can
be preserved by adding a boundary term to the bulk action, such that 
the supersymmetric variation of this boundary term 
exactly cancels the boundary piece generated by the 
supersymmetric transformation of the bulk action \cite{boundary}-\cite{boundary1}. 
This has been used for analysing the open M2-branes in the ABJM theory \cite{faizal}-\cite{faizal2}. This has also 
been used for studding 
 the BLG theory in presence of a boundary \cite{faizal1}-\cite{faizal4}. 

It may be noted that  the BLG theory in presence of a boundary can be 
made gauge invariant by adding extra boundary degrees of freedom to it \cite{faizal1}. Thus, the boundary BLG also 
has a gauge symmetry associated with it. So, it cannot be quantized without getting rid 
of these unphysical degrees of freedom. This 
can be done by adding   ghost and gauge fixing terms to the original classical Lagrangian.
The addition of these terms incorporates the gauge fixing condition at a quantum level. 
It is known that 
for  a gauge theory  the sum of the original classical Lagrangian with the gauge fixing and the ghost terms, 
is invariant under  new sets of transformations called 
the BRST and the anti-BRST  transformations \cite{brst}-\cite{nlbrst1}. For the ABJM theory, the 
  BRST  and the anti-BRST symmetries  have been studied in both linear and non-linear gauges \cite{abm}.

The infinitesimal  BRST and the infinitesimal  anti-BRST
transformations have been generalized 
to finite field dependent BRST (FFBRST) and finite field dependent anti-BRST 
(anti-FFBRST) transformations \cite{jm}-\cite{antifbrst}.
This is done 
by first  making the infinitesimal global parameter occurring in the 
BRST or the anti-BRST transformations depend on fields occurring in the theory. Then this 
field dependent parameter  is integrated to obtain the FFBRST and anti-FFBRST transformations. 
Even though, these finite transformations  are a symmetry of the quantum action, they are not a symmetry of the 
functional measure. They can thus be used to relate a theory 
in one gauge to the same theory in a different gauge \cite{ffb}-\cite{sudm3}. 
So, FFBRST transformations can be used to 
overcome  a problem that a 
theory suffers from in a particular gauge. 
 For example, the axial gauge even for ordinary gauge theories has  problamatic poles associates 
 with it. Furthermore,  in the Coulomb gauge the time-like propagator is not damped and so the time integral does not 
 converge. There is no reason not to expect similar problems to occur in non-covariant gauges in the BLG theory. 
 However, for ordinary gauge theories these problems can be resolved by using the FFBRST formalism. This is because  
 to overcome these type of difficulties calculation for the required quantity can be done for a 
gauge in which that problem does not exist, and then using the FFBRST transformation to 
transform it to the required gauge.
In fact, 
in ordinary Yang-Mills theory, FFBRST transformations  have been used for obtaining the propagator in Coulomb gauge 
  from the generating function  in the Lorentz gauge \cite{ff1}. Thus, FFBRST can be used for obtaining the propagators in 
  suitable gauges for the ABJM theory, which can be used for calculating scattering amplitudes in a suitable gauge.

  It may be noted that even though the BLG is a theory dual to supergravity theories, 
  and this duality  fixes the matter content of the theory, it is a gauge theory from 
  the field theoretic perspective. Thus, at a quantum level it can be analysed as a gauge theory using 
  the Gribov-Zwanziger theory. This way non-perturbative aspects of this theory can be analysed. 
There is a close relation between FFBRST transformation and Gribov-Zwanziger theory 
\cite{gz}. 
Thus, FFBRST transformations may give us an idea about the non-perturbative 
effects in a theory. 
It may be noted that the FFBRST symmetry relies crucially on the gauge sector of the theory, 
and will not  get effected by the matter content of the theory. Hence, even though 
the BLG theory has a very precise matter content fixed from supergravity theories, 
the FFBRST formalism can be applied to it. 
This is very important from the M-theory point of view. 
This is because we may be able to understand  the physics of multiple 
M5-branes by analysing non-perturbation effects in the BLG theory \cite{m5}-\cite{m512}.  
The FFBRST symmetry for the BLG theory on a manifold without a boundary  has already been  studied 
\cite{fs}. However, so far anti-FFBRST symmetry for BLG theory has not been studied. 
Furthermore,  as the M2-branes can end on M5-branes,
it is important to study the BLG theory in presence of a boundary. 
So, in this paper we will study 
 the anti-FFBRST transformations for the BLG theory in presence of a boundary.

\section{ BLG Theory in Presence of a Boundary }
In this section we will review the BLG theory in presence of a boundary \cite{faizal1}-\cite{faizal4}. 
As has been demonstrated in the previous section, we have  to add a boundary 
term to the conventional BLG theory to preserve half of the supersymmetry. 
In fact, we have to add yet another term to it to make it gauge invariant. 
However,  unlike regular gauge theories, where the gauge fields take values in a 
Lie algebra, the gauge field in the BLG theory take values in a Lie $3$-algebra. A Lie $3$-algebra is 
vector space endowed with a trilinear product, 
$
[T^A,T^B,T^C] = f^{ABC}_D T^D,
$
where  $T^A$ are called the generators of this Lie $3$-algebra. 
 These structure constants are totally antisymmetric 
  in $A,B,C$ and  satisfy the Jacobi identity, 
$
f^{[ABC}_G f^{D]EG}_H = 0
$.
It is also useful to define the following constants, 
$
C^{AB,CD}_{EF} = 2f^{AB[C}_{[E} \delta^{D]}_{F]} 
$ \cite{blgblg}. 
These constants are anti-symmetric in the  pair of indices $AB$ and $CD$ and the satisfy the following Jacobi identity,
$
C^{AB,CD}_{EF} C^{GH,EF}_{KL} + C^{GH,AB}_{EF} C^{CD,EF}_{KL} +C^{CD,GH}_{EF} C^{AB,GH}_{KL} =0 $.
Furthermore, the metric in the Lie $3$-algebra can be written as  $h_{AB} = Tr (T_A T_B) $.  

The BLG theory on manifolds without boundaries has $\mathcal{N}  =8$ supersymmetry. 
However, we will perform our analysis in $\mathcal{N} =1 $ superspace formalism where 
only the supersymmetry generated by $Q_a = \partial_a - (\gamma^\mu \partial_\mu \theta)_a$ is manifest. 
To write the BLG Lagrangian in presence of a boundary, we will first define covariant derivatives 
for  matter fields $X_A, X_A^{\dagger}$ and the spinor field $\Gamma^a_{AB}$ as \cite{blgblg}
\begin{eqnarray}
  \nabla_a  X^I_A&=& D_a X_A -i f^{BCD}_A \Gamma_{a BC} X^{I}_{ D},\nonumber\\
 \nabla_a X^{I \dagger}_A &=& D_a X^{I \dagger}_A + if^{BCD}_A X^{I \dagger}_D \Gamma^{a BC}, \\
( \nabla_a \Gamma_b)_{AB} &=& D_a \Gamma_{b AB} + C^{CD,EF}_{AB}\Gamma_{CD a} \Gamma_{b EF},  
\end{eqnarray}
where $D_a = \partial_a + (\gamma^\mu \partial_\mu)^b_a \theta_b$.
Under gauge transformations these fields transform as 
$
    \Gamma_{a } \rightarrow  i u \, \nabla_a u^{-1}, \, 
X^{I }\rightarrow  u X^I,\,
 X^{I  \dagger} \rightarrow X^{I \dagger} u^{-1},\, 
$
where 
$X^I = X^I_A T^A, \, X^{I \dagger} =  X^{I \dagger}_AT^A, \, 
 \Gamma_{a } = \Gamma_{a AB} T^A T^B $. We also define, 
 $\Gamma^a \times \Gamma_a = T^E T^F C^{AB,CD}_{EF}\Gamma^a_{AB} \Gamma_{a CD}$.
Now we define, 
\begin{eqnarray}
 \Omega_{ a AB} & = & \omega_{a AB} - \frac{1}{3}C^{CD,EF}_{AB}[\Gamma^{b CD}, \Gamma_{ab EF}] \\
 \omega_{a AB} & = & \frac{1}{2} D^b D_a \Gamma_{b AB} -i  C^{CD,EF}_{AB}[\Gamma^b_{CD} , D_b \Gamma_{a EF}] \nonumber \\ && -
 \frac{1}{3}C^{CD,EF}_{AB} C^{GH,IJ}_{EF}[ \Gamma^b_{CD} ,
\{ \Gamma_{b GH} , \Gamma_{a IJ}\} ],  \\
 \Gamma_{ab AB} & = & -\frac{i}{2} \left[ D_{(a}\Gamma_{b) AB} 
- 2 i C^{CD,EF}_{AB}\{\Gamma_{a CD}, \Gamma_{b EF}\} \right]. 
\end{eqnarray}
The Lagrangian for the BLG theory on a manifold without a boundary can  be written as \cite{fs}
\begin{eqnarray}
 \mathcal{L} &=&\nabla^2 \left [ \frac{k}{4\pi} f^{ABCD}\Gamma^{a}_{ AB}  \Omega_{a CD}
 +  \frac{1}{4} 
 (\nabla^a X^I)^A  (\nabla_a X^{\dagger}_I)_A \right. \nonumber \\  &&\left.  -\frac{2\pi}{k}
\epsilon_{IJKL} f^{ABCD}X^I_A X^{K \dagger}_B X^{J}_C  X^{L \dagger}_D\right]_{\theta =0}. 
\end{eqnarray}

Now we project 
the covariant derivatives $\nabla_{\pm b} = (P_{\pm})^a_b \nabla_a $ using the 
a projection operator, $(P_\pm)^b_a = (\delta^b_a \pm (\gamma^3)^b_a)/2$.
Under the action of this  projection operator $Q_a$ splits into $Q_{\pm b}$,
where $Q_{\pm b} = (P_{\pm})^a_b Q_a$ \cite{boundary}-\cite{boundary1}.  
 If we put a boundary at fixed $x_3$, then $\mu$ splits into $\mu = (\mu, 3)$, 
and  only the supersymmetry generated by  one of these supercharges can be preserved. 
The induced value of the matter and spinor fields on the boundary 
is denoted by $X_A', {X_A'}^{\dagger}$ and ${\Gamma^a_{AB}}'$, respectively.
Furthermore, the boundary covariant derivative of these induced fields, 
$\nabla_a'$, is obtained from $\nabla_a$, by neglecting $\gamma^3\partial_3$ contributions to it.
We also introduce a boundary superfield $v'$ and
let $v$ be its extension into the bulk with the following gauge transformation,  $v \rightarrow v u^{-1}$. 
The Lagrangian for the boundary BLG theory which is invariant under 
supersymmetry generated by $Q_+$ can be written as follows \cite{faizal1}-\cite{faizal4}
\begin{eqnarray}
 \mathcal{L}_{sg}&=& - \nabla_{+}' [\mathcal{CS}(\Gamma^v) + \mathcal{M} (X^{I }, X^{\dagger I })
 +\mathcal{K}' (\Gamma', v') ]_{\theta_- =0},
\end{eqnarray}
where $\Gamma^v_a$ denote the gauge transformation  of $\Gamma_a$ generated by $v$ and 
\begin{eqnarray}
\mathcal{CS} (\Gamma) &=&\frac{k}{4\pi} \nabla_-[f^{ABCD}\Gamma^{a}_{ AB}  \Omega_{a CD}]_{\theta_{+} =0}, 
\nonumber \\ 
\mathcal{M} ( X^I, X^{\dagger I }) &=&  \frac{1}{4} \nabla_- [
 (\nabla^a X^I)^A  (\nabla_a X^{\dagger}_I)_A ]_{\theta_{+} =0}\nonumber \\  && -\frac{2\pi}{k}\nabla_- [
\epsilon_{IJKL} f^{ABCD}X^I_A X^{K \dagger}_B X^{J}_C  X^{L \dagger}_D]_{\theta_{+} =0}, \nonumber \\
\mathcal{K}' (\Gamma', v' )&=&  -\frac{k}{2\pi} 
[ f_{ABCD}
({v}'^{-1} \nabla_{+}' v')^{AB} ({{v}'}^{-1} \mathcal{D}_{-}' v')^{CD}]_{\theta_{+} =0}. 
\end{eqnarray}
It may be noted that
 $\mathcal{S}(\Gamma', v')  =\mathcal{CS}(\Gamma^v) - \mathcal{CS}(\Gamma)$ is the boundary potential.  So, 
$\mathcal{CS}(\Gamma^v) =\mathcal{CS}(\Gamma) +  \mathcal{S}(\Gamma', v') ) $ is the total
potential of the theory. In case there is no coupling to the bulk fields this reduces to a potential term
 for the supersymmetric Wess-Zumino-Witten models,
\begin{eqnarray}
\nabla_{+}' \mathcal{S}(\Gamma', v') &=& -\frac{k}{2\pi} {\nabla}_{+} 'C^{CD, EF}_{AB}
\left[ [({v^{-1}}' \mathcal{D}_{-}' v')^{AB},  
({v^{-1}}' \mathcal{D}'_3 v')_{CD}]\right. \nonumber \\ && \left. \times
({v^{-1}}' \nabla_{+}' v')_{EF} \right]_{\theta_- =0}.
\end{eqnarray}

\section{Finite Field Dependent Transformation}
In this section we will construct  finite field dependent  anti-BRST transformations 
for the BLG theory in presence of a boundary. To do that we first note that
we can write the  gauge fixing term for the BLG theory as
\begin{equation}
\mathcal{L}_{gf} =\nabla_+ \nabla_- 
 \left[f^{ABCD}b_{AB}  D^a \Gamma_{a CD} + \frac{\alpha}{2}f^{ABCD} b_{AB}  b_{CD}  
\right]_{\theta =0}.
\end{equation}
The ghost term  corresponding to this gauge fixing term can be written as  
\begin{equation}
\mathcal{L}_{gh} = \nabla_+ \nabla_- 
\left[ f^{ABCD}\bar{c}_{AB}   D^a \nabla_a   c_{CD}
\right]_{\theta =0}.
\end{equation}
The total Lagrangian, $\mathcal{L}_{BLG}=\mathcal{L}_{sg}+\mathcal{L}_{gf}+\mathcal{L}_{gh}$, is now invariant under the following anti-BRST transformations, 
\begin{eqnarray}
\bar s \,\Gamma_{a} = \nabla_a   \bar c \ ,  && 
\bar s \,\bar c = - \bar c \times \bar c \   ,
\nonumber \\\bar s \, X^{ I \dagger }
 =  - i  X^{I \dagger } \bar c\ , 
 && 
\bar s \,b =- b \times \bar c\ ,  \nonumber \\ 
\bar s \, X^{I  }=i\bar c  X^{I }\ , &&   
\bar s\,{c} =- b- c\times c \ ,
   \nonumber \\
 \bar s\, v = -i v \bar c \ .&&  
\end{eqnarray}
In fact, it is also possible to write a non-linear parts of the above anti-BRST transformations as 
\begin{eqnarray}
\bar s \,{c} &=&- b\   - \frac{1}{2} {\bar c\times c } \   , 
 \nonumber \\ 
\bar s \,b&=&  - \frac{1}{2} {\bar  c\times b }  \    
+ \frac{1}{8}   \bar  c\times
\bar  c\times   c \   ,
 \nonumber \\ 
\bar s \,\bar c&=&  - \frac{1}{2} {\bar c\times\bar  c } \ .
\end{eqnarray}
Now the  infinitesimal global parameter 
    had to be included in the definition of  $\overline s$. 
However, we can explicitly write this global parameter with odd Grassmann parity
occurring in these transformations  as $\epsilon$.  
 The properties of the anti-BRST 
transformation do not depend on whether this parameter is
field dependent or not.
Similarly, they do not depend on whether this parameter  is  finite or infinitesimal. 
Thus, we first   make this infinitesimal global 
parameter with odd Grassmann parity  an infinitesimal field dependent parameter with odd Grassmann parity
$\epsilon [{\Phi}   (x, \kappa)]$, 
where $\Phi^{i } (x, \kappa)
 = (X^I (x, \kappa), X^{ I \dagger }(x, \kappa), \Gamma_a(x, \kappa), c(x, \kappa), \bar{c}(x, \kappa), 
 b(x, \kappa), v(x, \kappa))$. 
  Here   ${\Phi^i}   (x, 0 )$ are the initial fields and
 $ {\Phi^i}   (x, 1)$ are the transformed field.
Furthermore,  $\kappa: 0\le \kappa \le 1$ is an arbitrary parameter and physical 
quantities do not depend on it  \cite{cdec, jm}.

Now we can make this infinitesimal field dependent parameter a  finite field dependent parameter and thus define a 
functional with odd Grassmann parity as $\Theta [{\Phi}   ]$. This finite field dependent parameter  can  obtained from a
infinitesimal
field dependent parameter  as follows, 
\begin{equation}
\Theta  [{\Phi}   (x, \kappa)] = \epsilon  [{\Phi}   (x, \kappa)] \frac{ \exp F [{\Phi}   (x, \kappa)]
-1}{F [{\Phi}   (x, \kappa)]},
\end{equation} 
where 
\begin{eqnarray}
 F &=& \frac{ \delta \epsilon [\Phi (x, \kappa)]}{\delta
X^I (x, \kappa)} \bar s  X^I (x, \kappa) + \frac{ \delta \epsilon [\Phi (x, \kappa)]}{\delta
X^{I\dag} (x, \kappa)} \bar s  X^{I\dag} (x, \kappa) +  \frac{ \delta \epsilon [\Phi (x, \kappa)]}{\delta
\Gamma_a (x, \kappa)} \bar s  \Gamma_a (x, \kappa)\nonumber \\ && +
\frac{ \delta \epsilon [\Phi (x, \kappa)]}{\delta
c(x, \kappa)} \bar s  c (x, \kappa)   +
\frac{ \delta \epsilon [\Phi (x, \kappa)]}{\delta
\bar c (x, \kappa)}\bar  s  \bar c (x, \kappa)+
\frac{ \delta \epsilon [\Phi (x, \kappa)]}{\delta
b(x, \kappa)}\bar  s b(x, \kappa) \nonumber \\ && +
\frac{ \delta \epsilon [\Phi (x, \kappa)]}{\delta
v(x, \kappa)}\bar  s v(x, \kappa) . \label{ref}
\end{eqnarray}
For a field dependent  infinitesimal parameter $\epsilon [{\Phi}   (x,\kappa )]$, we have 
\begin{equation}
\frac{ d}{d \kappa}{\Phi^i}   (x, \kappa ) = \bar s  {\Phi^i}    (x, \kappa )\
\epsilon [{\Phi}   (x,\kappa )].
\label{dif}
\end{equation}
We  integrate  these equations 
from $ \kappa=0$ to $\kappa=1$ and  find the relation between 
 ${\Phi^i} ( x, 1) $ and  ${\Phi^i}   (x, 0)
$, 
\begin{equation}
{\Phi^i} (x, 1) = {\Phi^i}   (x, 0) + \bar s  {\Phi^i}    (x, 0) \Theta [{\Phi}   (x)],
\end{equation}
Now the anti-FFBRST transformations for the BLG theory are given by
$
f\,\Phi^{i }  = \bar s \,  \Phi^{i } \Theta 
$. 
The anti-FFBRST transformations are a symmetry of the action $S_{BLG}$. However, they are  not a symmetry 
of the functional measure
because the Jacobian for path integral measure  in the expression of generating functional  is not invariant
under them. 
The change in the Jacobian under the  FFBRST transformation
is given by 
$
{\cal D}{\Phi^i}    =J[{\Phi}   (\kappa)] {\cal D}{\Phi^i}   (\kappa). \label{jac}
$            
Now $J[{\Phi}   (\kappa)]$ can be replaced within the 
functional integral by 
$e^{iS_1[{\Phi}   (\kappa)]}$, where  $S_1[{\Phi}   (\kappa)]$ is some 
local functional of ${\Phi^i}$. The condition
 for existence of $S_1$ for the BLG theory in presence of a boundary is given by 
\begin{eqnarray}
\int d^3x \nabla_+ \nabla_-  \left[\frac{1}{J (\kappa )}\frac{d J (\kappa )}{d\kappa} -i\frac{dS_1}{d\kappa}\right]_{\theta =0} =0.
\label{mcond}
\end{eqnarray}
The 
infinitesimal change in Jacobian is given by 
\begin{equation}
 \delta \mathcal{J} (\kappa) = \nabla_+ \nabla_-  \left[ \frac{1}{J(\kappa)}\frac{dJ(\kappa)}{d\kappa}\right]_{\theta =0} .
\end{equation}
So, we can write 
 \begin{eqnarray} 
 \delta \mathcal{J} (\kappa) &=& 
 -\int d^3x \nabla_+ \nabla_-     \left[- \bar s  X^I (x, k) \frac{  \delta \epsilon [\Phi (x, k)]}{\delta
X^I (x, k)} \right. \nonumber \\ &&\left. - \bar s  X^{I\dag} (x, k)\frac{ \delta \epsilon [\Phi (x, k)]}{\delta
X^{I\dag} (x, k)}    + \bar s  \Gamma_a (x, k) \frac{ \delta \epsilon [\Phi (x, k)]}{\delta
\Gamma_a (x, k)} \right. \nonumber \\ &&\left. - \bar s  c (x, k)
\frac{ \delta \epsilon [\Phi (x, k)]}{\delta
c(x, k)}    - \bar s  \bar c (x, k)
\frac{ \delta \epsilon [\Phi (x, k)]}{\delta
\bar c (x, k)}  \right. \nonumber \\ &&\left. +\bar s b(x, k)
\frac{ \delta \epsilon [\Phi (x, k)]}{\delta
b(x, k)}
+ \bar s  v (x, k)
\frac{ \delta \epsilon [\Phi (x, k)]}{\delta
 v (x, k)} 
\right]_{\theta =0}.\label{jaceva}
\end{eqnarray}

\section{Anti-FFBRST Transformations with specific parameter}
We will now use anti-FFBRST to relate the BLG theory, in the linear gauge to it in the non-linear gauge. 
Here the linear anti-BRST transformations are denoted by $(\bar sG_1)^{AB}$ and the 
the non-linear anti-BRST transformations are denoted by $(\bar sG_2)^{AB}$. 
We again  define the infinitesimal field dependent parameter as follows  
\begin{eqnarray}
\epsilon [\Phi] = 
-i\gamma\int d^3x \nabla_+ \nabla_-    \left[ f^{ABCD} c_{AB} \left( G_{CD 1} - G_{CD 2}\right)\right]_{\theta =0}, 
\end{eqnarray}
where $\gamma$ is an arbitrary constant parameter.
However, the expression for the change in Jacobian is now given by 
\begin{eqnarray}
 \delta \mathcal{J} (\kappa)
&=&i\gamma\int d^3x \nabla_+ \nabla_-    f^{ABCD}\left[ ( \bar s G_{CD 1} - \bar sG_{CD 2})c_{AB}
\right.
\nonumber\\
 &&\left.
+ (b_{AB} +C_{AB}^{GH,EF}\bar c_{GH}c_{EF}) (G_{CD 1} - G_{CD 2})\right]_{\theta =0}.
\end{eqnarray}
Furthermore, we construct the local functional as
\begin{eqnarray}
S_1 &=&\int d^3 x \nabla_+ \nabla_-   f^{ABCD}[\xi_1 (\kappa) b_{AB}G_{CD 1}+  \xi_2 (\kappa) b_{AB}G_{CD 2} \nonumber \\ 
& &+\xi_3(\kappa)  
\bar s G_{CD 1} c_{AB}+\xi_4(\kappa) \bar sG_{CD 2}  c_{AB } \nonumber \\ 
& & +\xi_5(\kappa)C_{AB}^{GH,EF}\bar
c_{GH}c_{EF}G_{CD1}\nonumber\\
&&+\xi_6(\kappa)C_{AB}^{GH,EF}\bar
c_{GH}c_{EF}G_{CD2}]_{\theta =0},
\end{eqnarray}
where $\xi_i(i=1,2,3,4)$ are the $\kappa$ dependent 
arbitrary parameters which satisfy the following initial boundary condition
$
\xi_i (\kappa =0)=0
$. 
As all the fields again depend on $\kappa$, so we calculate
\begin{eqnarray}
\frac{dS_1}{d\kappa}&=&\int  d^3x \nabla_+ \nabla_-   f^{ABCD}[\xi'_1  b_{AB}G_{CD 1} 
-\xi_1 C_{AB}^{GH, EF} b_{GH}\bar c_{EF} G_{CD1}\epsilon \nonumber\\
&&+\xi_1 b_{AB}\bar s G_{CD1}\epsilon +  \xi'_2  b_{AB}G_{CD 2}-\xi_2 C_{AB}^{GH, EF} b_{GH}\bar c_{EF} G_{CD2}\epsilon
\nonumber\\ & &
+\xi'_3 
\bar s G_{CD 1} c_{AB} -\xi_3  
\bar s G_{CD 1}(b_{AB} +C_{AB}^{GH, EF}\bar c_{GH}c_{EF})\epsilon  \nonumber\\
&&+\xi'_4  \bar sG_{CD 2}  c_{AB }
  -\xi_4 \bar s G_{CD 2}(b_{AB} +C_{AB}^{GH, EF}\bar c_{GH}c_{EF})\epsilon\nonumber\\
  &&+\xi_5' C_{AB}^{GH,EF}\bar
c_{GH}c_{EF}G_{CD1}+\xi_6' C_{AB}^{GH,EF}\bar
c_{GH}c_{EF}G_{CD2}\nonumber\\  
&&-\xi_5 C_{AB}^{GH,EF}\bar c_{GH}b_{EF}G_{CD1}\epsilon +\xi_5
C_{AB}^{GH,EF}\bar c_{GH}c_{EF}\bar sG_{CD1}\epsilon \nonumber\\
&&-\xi_6 C_{AB}^{GH,EF}\bar c_{GH}b_{EF}G_{CD2}\epsilon +\xi_6
C_{AB}^{GH,EF}\bar c_{GH}c_{EF}\bar sG_{CD2}\epsilon \nonumber \\ &&
+\xi_2 b_{AB}\bar s G_{CD2}\epsilon ]_{\theta =0},\nonumber\\
 &=&\int  d^3x \nabla_+ \nabla_-  f^{ABCD}[\xi'_1  b_{AB}G_{CD 1} +\xi'_3 \bar s G_{CD 1} c_{AB} 
\nonumber \\ &  & +\xi_5' C_{AB}^{GH,EF}\bar
c_{GH}c_{EF}G_{CD1}+\xi_6' C_{AB}^{GH,EF}\bar
c_{GH}c_{EF}G_{CD2}\nonumber\\
&&+(\xi_1 -\xi_3 )\bar s G_{CD 1}b_{AB}\epsilon  -(\xi_3 -\xi_5) C_{AB}^{GH, EF}\bar 
c_{GH}c_{EF}\bar s G_{CD1}\nonumber \\ & & +(\xi_2 -\xi_4)\bar s G_{CD 2} b_{AB} 
\epsilon   -(\xi_4 -\xi_6) C_{AB}^{GH, EF}\bar c_{GH}c_{EF}\bar s G_{CD2}\epsilon\nonumber\\
&&-(\xi_1 -\xi_5) C_{AB}^{GH, EF} b_{GH}\bar c_{EF} G_{CD1}\epsilon +\xi'_4\bar s G_{CD 2} c_{AB}\nonumber\\
&&-(\xi_2 -\xi_6) 
C_{AB}^{GH, EF} b_{GH}\bar c_{EF} G_{CD2}\epsilon   +  \xi'_2  b_{AB}G_{CS 2}
]_{\theta =0}.
\end{eqnarray}
Now, the Jacobian can be written as  $e^{iS_1}$, if the following equation is satisfied 
\begin{eqnarray}
&&\int d^3x \nabla_+ \nabla_-  [ 
 f^{ABCD}[(\xi'_1 -\gamma)  b_{AB}G_{CD 1} +  (\xi'_2 +\gamma)  b_{AB}G_{CS 2}
\nonumber \\ &  & +(\xi_5' -\gamma) C_{AB}^{GH,EF}\bar
c_{GH}c_{EF}G_{CD1}+(\xi_6' +\gamma) C_{AB}^{GH,EF}\bar
c_{GH}c_{EF}G_{CD2}\nonumber\\
&&+(\xi_1 -\xi_3 )\bar s G_{CD 1}b_{AB}\epsilon  -(\xi_3 -\xi_5) C_{AB}^{GH, EF}\bar 
c_{GH}c_{EF}\bar s G_{CD1}\nonumber \\ & &+(\xi_2 -\xi_4)\bar s G_{CD 2} b_{AB} 
\epsilon   -(\xi_4 -\xi_6) C_{AB}^{GH, EF}\bar c_{GH}c_{EF}\bar s G_{CD2}\epsilon\nonumber\\
&&-(\xi_1 -\xi_5) C_{AB}^{GH, EF} b_{GH}\bar c_{EF} G_{CD1}\epsilon +(\xi'_4 +\gamma )\bar s G_{CD 2} c_{AB}\nonumber\\
&&-(\xi_2 -\xi_6) 
C_{AB}^{GH, EF} b_{GH}\bar c_{EF} G_{CD2}\epsilon \nonumber\\
&&
+(\xi'_3
-\gamma ) \bar s G_{CD 1} 
c_{AB} ]_{\theta =0}=0.
\end{eqnarray}

So,  equating the coefficients of the above expression,  we get 
$
 \xi'_1 -\gamma =0,  \, \xi_2' +\gamma =0,\, \xi'_3 -\gamma =0,\, \xi'_4 +\gamma =0, \, 
\xi_5 -\gamma =  0,\, \xi_6 +\gamma =0,
$ and 
$\xi_1 -\xi_3 =\xi_2 -\xi_4=\xi_3 -\xi_5=\xi_4 -\xi_6=\xi_1 -\xi_5=\xi_2 -\xi_6 =0 $.
For $\gamma =-1$, the solutions to above equations satisfying initial condition for $\xi$  are
given by 
$
\xi_1 = -\kappa, \, \xi_2 = \kappa, \, \xi_3 = -\kappa, \, \xi_4 = \kappa,\ \xi_5 =-\kappa,\
\xi_6 =\kappa
$.
Now, by adding $S_1 (\kappa =1 )$ to the original action in 
the gauge $G_{CD 1}$, we obtain the action in gauge $G_{CD 2}$,  $S_{f} = S_{BLG } + S_1$.
So,   under the anti-FFBRST transformations the  generating functional 
in the gauge $G_{CD 2}$ transforms to the generating functional 
 in the gauge $G_{CD 1}$. In fact,  the anti-FFBRST transformations can also be used to obtain the generating functional 
in the gauge $G_{CD 2}$, if we start from the generating functional in the gauge $G_{CD 1}$.

\section{Conclusion}

In this paper we analysed the BLG theory on a manifold with a boundary. This theory was made gauge invariant 
by adding a new field to it, such that the boundary term  generated by the gauge transformations of the 
BLG theory exactly canceled the boundary term  generated by the gauge transformations of this new term. 
The measure of integration of the superspace was also modified so that the theory preserves half the supersymmetry even 
in presence of a boundary. Furthermore,  the anti-BRST transformations of this theory were studied 
in both linear as well as non-linear gauges. 
After analysing the both the linear and non-linear  anti-BRST transformations, the anti-FFBRST 
transformations were constructed. It was demonstrated that these two gauges can be related to each other via anti-FFBRST 
transformations. 

The main motivation for studding the FFBRST or the  anti-FFBRST transformation of the BLG theory is that they 
 are related to  Gribov-Zwanziger theory and can thus give us an idea about the non-perturbative 
effects in the BLG theory. As the physics of multiple 
M5-branes can be studied using non-perturbation effects in the BLG theory \cite{m5}-\cite{m512}, 
FFBRST transformations for the BLG theory are very important.
Furthermore, as M2-branes can end on M5-branes, it is important to study the
FFBRST transformations for the 
BLG theory 
on a manifold with boundaries. 
However, it may be noted that the
BLG theory only describes two M2-branes. It has been generalized to the ABJM theory, which   describes 
more than two M2-branes \cite{abjm}. 
The BRST and the anti-BRST symmetries for the ABJM theory have been studied in various gauges \cite{abm}. 
 Thus, it would be interesting to study the FFBRST and the anti-FFBRST transformations 
for the ABJM theory. 
The ABJM theory in presence of a boundary has also been constructed \cite{faizal}-\cite{faizal2}.
In fact, the BRST and the anti-BRST transformations for the ABJM theory in presence of a boundary have also 
been discussed \cite{faizal}.  So, it would also be interesting to find a finite field version of these 
transformations. However, as
the ABJM theory has two infinitesimal gauge parameters, the conventional FFBRST transformation 
would have to be modified for the ABJM theory. It may be noted that the ABJM theory coincides with the BLG theory
for two M2-branes. So, for two M2-branes,
the FFBRST transformations for the ABJM theory should reduce to the FFBRST transformations for the BLG theory. 
The FFBRST transformation for gauge symmetry
generated by a Lie $3$-algebra can used for  analysing the FFBRST transformation
for the gauge theory with two parameters as is possible to 
decompose  $SO(4)$ into $SU (2) \times SU (2)$. 
This can then be used to motivate the analysis of FFBRST transformations for the ABJM theory. 

It may be noted that multiple D2-brane action can be derived from a multiple M2-brane action by 
means of a novel Higgs mechanism \cite{d2}-\cite{d21}. In this mechanism 
a vacuum expectation value is given to a scalar field. This breaks  the gauge group of the ABJM theory
 down to its diagonal subgroup. Thus, the Yang-Mills theory coupled to matter fields is obtained
 by Higgsing the ABJM theory. 
It would be interesting to start with a gauge fixed ABJM theory and derive its FFBRST transformations, 
then using the novel Higgs mechanism to go to the multiple D2-branes action. 
It would be expected that the FFBRST transformations 
in this case will reduce to the FFBRST transformations for the D2-branes. By analysing these transformations, 
we can get a better understanding of what happens to this theory at a quantum level. We can also 
perform a similar analysis using anti-FFBRST transformations. 

 It may be noted   that for Yang-Mills theory  in Cho-Faddeev-Niemi variables
 with an appropriate choices of finite and field dependent parameter
the gauge fixing and ghost terms corresponding to Landau gauge and  maximal Abelian gauge 
appear naturally  by using the FFBRST transformations \cite{fbaa}. It will be interesting to 
perform a similar analysis for M2-branes and show that 
the gauge fixing and ghost terms corresponding to different gauges can also
occur naturally for M2-branes  by using the FFBRST transformations.  
It may be noted that various deformations of the general relativity  have been studied \cite{H1}-\cite{H}, 
and it would be possible to study such deformations of supergravity 
in eleven dimensions. 
So, it will be interesting to analyse the action for  M2-branes dual to such deformations of supergravity in eleven dimensions, 
and repeat the analysis of this paper for such deformations.

\end{document}